\documentclass[useAMS,usenatbib]{mn2e}

\IfFileExists{srcltx.sty}{\usepackage[active]{srcltx}}

\usepackage{graphicx,xspace}

\newcommand{\dm}{{\textsc{dm}}} 
\newcommand{\kev}{\ensuremath{\:\mathrm{keV}}} %

\newcommand{\wdm}{{\rm wdm}} %

\newcommand{\hMsun}{$\rm {M_{\odot}h^{-1}}$}
\newcommand{\etal}{et al.~}
\newcommand{\rhocrit}{\rho_c}
\newcommand{\rhorms}{\rho_{\rm rms}}

\newcommand{\cvir}{c_{\rm vir}}
\newcommand{\Rvir}{R_{\rm vir}}
\newcommand{\Mvir}{M_{\rm vir}}
\newcommand{\rs}{r_{\rm s}}
\newcommand{\Jvir}{J_{\rm vir}}
\newcommand{\Vvir}{V_{\rm vir}}
\newcommand{\Nvir}{N_{\rm vir}}

\newcommand{\m}[1]{\textbf{m#1}\xspace} %

\newcommand{\mnras}{MNRAS}
\newcommand{\apj}{ApJ}
\newcommand{\apjl}{ApJ}

\def \kms {\ifmmode  \,\rm km\,s^{-1} \else $\,\rm km\,s^{-1}  $ \fi }
\def \kpc {\ifmmode  {\rm kpc}  \else ${\rm  kpc}$ \fi  }  
\def \Msun {\ifmmode M_{\odot} \else $M_{\odot}$ \fi} 
\def \hMsun {\ifmmode h^{-1}\,\rm M_{\odot} \else $h^{-1}\,\rm M_{\odot}$ \fi}

\def \rhos {\ifmmode \rho_{\rm s} \else $\rho_{\rm s}$ \fi} 
\def \rs {\ifmmode r_{\rm s} \else $r_{\rm s}$ \fi} 
\def \cvir {\ifmmode c_{\rm vir} \else $c_{\rm vir}$ \fi} 
\def \Rvir {\ifmmode r_{\rm vir} \else $R_{\rm vir}$ \fi}
\def \Vvir {\ifmmode V_{\rm  vir} \else  $V_{\rm vir}$  \fi} 
\def \Mvir {\ifmmode M_{\rm  vir} \else $M_{\rm  vir}$ \fi}  
\def \Nvir {\ifmmode N_{\rm  vir} \else $N_{\rm  vir}$ \fi}  
\def \Jvir {\ifmmode J_{\rm vir} \else $J_{\rm vir}$ \fi} 
\def \Evir {\ifmmode E_{\rm vir} \else $E_{\rm vir}$ \fi} 
\def \lam {\ifmmode \lambda  \else $\lambda$ \fi} 
\def \lamp {\ifmmode \lambda^{\prime} \else $\lambda^{\prime}$  \fi} 
\def \Vmax {\ifmmode V_{\rm  max} \else  $V_{\rm max}$  \fi}

\title[The inner structure of haloes in CWDM] {The inner structure of haloes
  in Cold+Warm dark matter models} \author[ A. V. Macci{\`o} \etal] {Andrea
  V. Macci\`o$^{1}$\thanks{email: maccio@mpia.de},
  Oleg Ruchayskiy$^2$, Alexey Boyarsky$^{3,4}$, Juan C. Mu\~noz-Cuartas$^5$\\
  $^1$ Max-Planck-Institut f\"{u}r Astronomie, K\"{o}ningstuhl 17, 69117
  Heidelberg, Germany.\\
  $^2$ CERN Physics Department, Theory Division, CH-1211 Geneva 23,
  Switzerland\\
  $^3$ Instituut-Lorentz for Theoretical Physics, Universiteit Leiden,
  Niels Bohrweg 2, Leiden, The Netherlands\\
  $^4$ Bogolyubov Institute of Theoretical Physics, Kyiv, Ukraine\\
  $^5$ Leibniz-Institut f\"ur Astrophysik Potsdam, An der Sternwarte 16, 14482 Potsdam, Germany}

\begin{document}

\date{Accepted XXXX, Received XXXX}

\pagerange{\pageref{firstpage}--\pageref{lastpage}} \pubyear{2011}

\maketitle

\label{firstpage}

\begin{abstract}
  We analyze the properties of dark matter halos in the cold-plus-warm dark
  matter cosmologies (CWDM).  We study their dependence on the fraction and
  velocity dispersion of the warm particle, keeping the free-streaming scale
  fixed.  To this end we consider three models with the same free-streaming:
  (1) a mixture of 90\% of CDM and 10\% of WDM with the mass 1 keV; (2) a
  mixture of 50\% of CDM and 50\% of WDM with the mass 5 keV; and (3) pure WDM
  with the mass 10 keV. ``Warm'' particles have rescaled Fermi-Dirac spectrum
  of primordial velocities (as non-resonantly produced sterile neutrinos would
  have). We compare the properties of halos among these models and with a
  $\Lambda$CDM with the same cosmological parameters. We demonstrate, that
  although these models have the same free-streaming length and the
  suppression of matter spectra are similar at scales probed by the
  Lyman-$\alpha$ forest (comoving wave-numbers $k < 3 - 5$ h/Mpc), the
  resulting properties of halos with masses below $\sim 10^{11}M_\odot$ are
  different due to the different behaviour of matter power spectra at smaller
  scales. In particular, we find that while the number of galaxies remains the
  same as in $\Lambda$CDM case, their density profiles become much less
  concentrated, and hence in better agreement with current observational constraints.
  Our results imply that a single parameter (e.g. free
  streaming length) description of these models is not enough to fully capture
  their effects on the structure formation process.
\end{abstract}



\section{Introduction}

It is usually said that cosmological data favour \emph{Cold} Dark Matter. The
cosmological ``concordance model'' is therefore often called
$\Lambda$CDM. However, a more precise statement should be that \emph{hot} dark
matter particles (i.e. the particles that became non-relativistic only around
recombination time, such as e.g. the ordinary neutrinos) are ruled
out~\citep{Davis:85}. The difference between cold and \emph{warm} DM particles
(the former being always non-relativistic and the latter becoming
non-relativistic deeply in the radiation-dominated epoch) would show up at
approximately galactic scales and it is only recently that such small scale
effects are starting to be resolvable both theoretically~\citep[see e.g.][and
refs. therein]{Maccio:09,Polisensky:10,Markovic:10,Semboloni:11,%
  vanDaalen:11,Lovell:11,Viel:11,Smith:11,Schneider:11,Dunstan:11} and
experimentally~\citep[see
e.g.][]{Viel:05,Viel:06,Seljak:06,Viel:07,Boyarsky:08c,Tikhonov:2009jq,%
  Zavala:2009ms,Song:2009sx,Papastergis:2011xe}. Warm dark matter N-body
simulations require significantly larger number of particles to resolve the
same scales as compared with the CDM case~(see
e.g.~\citealt{Wang:07,Lovell:11}). Additionally, at sub-Mpc scales baryonic
physics can hide (or mimic) the WDM suppression of power~(see
e.g.~\citep{Benson:01b,Bullock:00,Semboloni:11}, which makes the analysis of
small-scale data challenging.

The tiny (from the cosmological point of view) difference between cold and
warm dark matter is however of crucial importance for particle physics, as it
means a huge difference in the properties of corresponding particles and may
eventually provide a clue on the structure of a fundamental theory of
particles and interactions.

Historically the first WDM models were \emph{thermal relics} -- particles that
were in equilibrium in the early Universe and froze-out, being
relativistic~\citep{Colin:00,Bode:00}.  Such particles had thermal primordial
velocity spectrum and strong cutoff-like suppression of the power
spectra~\citep{Bode:00,Viel:05,Boyarsky:08c} at scales below few Mpc. Such
models are characterized by only one scale --- position of the cut-off in the
power spectrum, related to their free-streaming horizon.  One possible tool to
probe the growth of structures of (sub)Mpc scales is the Lyman-$\alpha$ forest
method -- studies of statistics of absorption features in the spectra of
distant quasars. The Lyman-$\alpha$ forest
data~\citep{Hansen:01,Viel:05,Viel:06,Seljak:06,Viel:07,Boyarsky:08c} puts
such strong constraints at their free-streaming length that ``thermal relics''
WDM models, compatible with Lyman-$\alpha$ bounds, produce essentially no
observable changes in the Galactic 
structures~\citep[c.f.][]{Strigari:06,Colin:07,Boyarsky:08c,deNaray:2009xj,Schneider:11}.

However, particle physics motivated WDM candidates can be produced in the
early Universe in non-thermal ways, may have significant non-zero primordial
velocities in the radiation-dominated epoch and non-equilibrium velocity
spectra~\citep[for review see
e.g.][]{Boyarsky:08d,Boyarsky:09a,Taoso:07,Feng:10}. In many models (e.g.\
sterile neutrinos, gravitino, axino) the same DM particles can be produced via
two co-existing mechanisms and therefore generically primordial velocity
spectra have ``colder'' and ``warmer'' components. Such models can be called
\emph{mixed} or ``cold plus warm'' dark matter models
(\textbf{CWDM})~(\citealt{Boyarsky:08c}, see also
\citealt{Palazzo:07}). Qualitatively, structures form in these models in a
bottom-up fashion (similar to CDM).  The way the scales are suppressed in CWDM
models is more complicated (and in general less severe for the same masses of
WDM particles), as comparable with pure warm DM models.  The first results
of~\citet{Lovell:11} demonstrate that the resonantly produced sterile neutrino
DM models, compatible with the Lyman-$\alpha$ bounds of~\citet{Boyarsky:08d},
do change the number of substructure of a Galaxy-size halo and their
properties.  The discrepancy between the number of observed substructures with
small masses and those predicted by $\Lambda$CDM models (first pointed out
in~\citealt{Klypin:99,Moore:99b}) can simply mean that these substructures did
not confine gas and are therefore completely dark \citep[see e.g.][]{Bullock:00,Benson:01b,Somerville:02,Maccio:10}.
This is not true for larger objects.  In particular, CDM numerical simulations
invariably predict several satellites ``too big'' to be masked by galaxy
formation processes, in contradiction with
observations~\citep{BoylanKolchin:11}.  Sterile neutrino DM of the minimal
neutrino extension of the Standard Model (the $\nu$MSM)~\cite{Asaka:05b,
  Boyarsky:09a}, with its non-trivial velocity dispersion, turns out to be
``warm enough'' to amend these issues~\citep{Lovell:11} and ``cold enough'' to
be in agreement with Lyman-$\alpha$ bounds~\citep{Boyarsky:08d}.

In this paper we study the structures of halos in the CWDM models.  To this
end we pick two CWDM models, compatible with the Lyman-$\alpha$ data of
\citet{Boyarsky:08c} and having the same free-streaming of the WDM particles.
We demonstrate that the properties of halos differ in these models (and differ
from both pure CDM and pure WDM model with the same free-streaming), meaning
that they are not determined by the free-streaming alone.

The paper is organized as follows. We discuss the choice of parameters of our
DM models and initial conditions for simulations in Section~\ref{ssec:IC}. The
suite of simulations is discussed in Sec.~\ref{sec:sims} and the main results
in Section~\ref{sec:results}. We discuss our results in
Section~\ref{sec:summ}.

\begin{figure}
  \centering
  \includegraphics[width=\linewidth, angle=270]{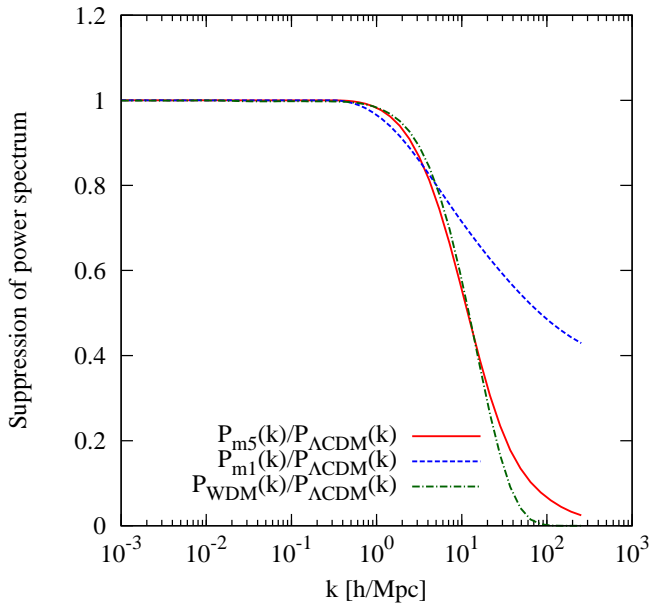}
  \caption{Ratio of the power spectra of the three models used in this work to
    that of $\Lambda$CDM (for the same values of cosmological parameters).
    The blue short-dashed curve is a mixture of 90\% of cold dark matter and
    10\% warm dark matter with rescaled Fermi-Dirac spectrum and mass
    $m_\textsc{wdm} = 1$ keV, (model \textbf{m1}); the red solid line is a
    mixture 50\% of cold and warm particles with the mass $m_\textsc{wdm} =
    5$~keV (model \textbf{m5}) and the green dashed-dotted line (model
    \textbf{wdm}) is a 100\% WDM model with the mass of DM particle
    $m_\textsc{wdm} = 10$ keV.  All three models are compatible with the
    Lyman-$\alpha$ analysis of~\citet{Boyarsky:08c}. }
  \label{fig:tf}
\end{figure}

\section{Model selection and initial conditions}
\label{ssec:IC}

For our simulations we selected two CWDM, one WDM and one reference $\Lambda$CDM model.
Our first CWDM
model (model \textbf{m1} in what follows) is a mixture of 90\% of cold dark
matter and 10\% of warm dark matter with the mass $m_{\wdm} = 1$ keV, while
the second is a 50\% cold and warm mixture with the mass $m_{\wdm} = 5$~keV
(model \textbf{m5} hereafter).  The WDM model has $m_{\wdm} = 10$~keV. In all
cases the WDM components have the rescaled Fermi-Dirac spectrum of primordial
velocities
\begin{equation}
  \label{eq:1}
  f(v) = \frac{\chi}{\exp\Bigl(\frac{m_{\wdm}v}{T_\nu(z)}\Bigr) + 1}
\end{equation}
where $m_{\wdm}$ is the WDM mass, $T_\nu(z)$ is the temperature of cosmic
neutrino background, evolving with redshift as $T_\nu(z) = T_{\nu0}(1+z)$,
$T_{\nu0} =1.9$~K, and the constant $\chi$ is determined by the requirement to
provide a given fraction, $F_\wdm$, of the total dark matter density:
\begin{equation}
  \label{eq:2}
  m^4_{\wdm} \int \frac{d^3v}{(2\pi)^3} f(v)  = F_\wdm \rho_\dm
\end{equation}
(in the units where $c=\hbar = k_B = 1$). Non-resonantly produced sterile
neutrinos would have such a phase-space
distribution~(\ref{eq:1})~\citep{Dodelson:93,Dolgov:00}.  One can relate the
parameters of such WDM models and those of ``thermal relics'' \citep{Bode:00},
using e.g. formulas in~\citet{Viel:05}.  In order to investigate the
dependence of the properties of the cosmologies on parameters other than
free-streaming (i.e. WDM mass and $F_\wdm$ component), we have chosen the
models in such a way that their \emph{free-streaming horizon is the same}.
Both models are compatible with the CWDM Lyman-$\alpha$ analysis
of~\cite{Boyarsky:08c}.

We computed the \emph{linear} power spectrum at redshift $z_{ini}\sim 30$ of
matter density perturbations $P_{ini}(k,z_{ini})$ for these models. The
standard software (i.e.  \textsc{camb},~\citealt{Lewis:99}) is not immediately
appropriate for this purpose, as it only treats massive neutrinos with a
Fermi-Dirac primordial distribution. To adapt it to the problem at hand, we
modified \textsc{camb} so that it could take arbitrary spectra as input data
files.  We analyzed the spacing in momentum space needed in order to obtain
precise enough results, and implemented explicit computations of distribution
momenta in \textsc{camb}.  We cross-checked our results by modifying another
linear Boltzmann solver -- \textsc{cmbfast}~\citep{Seljak:96}, implementing a
treatment of massive neutrinos with arbitrary \emph{analytic} distribution
function and with the \textsc{class} code~\citep{Lesgourgues:2011rh} where
such an option is realized.

Figure~\ref{fig:tf} shows the ratio of the power spectrum in our CWDM models
with respect to the $\Lambda$CDM model with the same cosmological
parameters. The effect of the free-streaming is clearly visible on scales
smaller than $k \approx 1$ h/Mpc.  It is important to notice that although the
two models formally have the same free-streaming length, since the product of
the WDM mass and its abundance is constant, the suppression of power on small
scales is different in the two models and larger for the \m5 model, due to its
more abundant warm dark matter component.

Based on the linear power spectrum $P_{ini}(k,z_{ini})$ the initial conditions
for N-body simulations are generated with a modified version of the GRAFIC2
package (Bertschinger 2001).  In this modified version the transfer function
at the starting redshift is given as an external input.  The starting
redshifts $z_i$ are set to the time when the standard deviation of the
smallest density fluctuations resolved within the simulation box reaches
$0.15$ (the smallest scale resolved within the initial conditions is defined
as twice the intra-particle distance).  We used the best-fit cosmological
parameters from~\citet{Boyarsky:08c}, comparable with WMAP5
results~\citep{WMAP5cosmoParams}.  Namely: $\Omega_m = 0.253$, $\Omega_\Lambda
= 0.747$, $n=0.973$, $h = 0.72$, and $\sigma_8 = 0.8$.

In our small box simulations ($L=20$ h$^{-1}$ Mpc ) we also include the effect
of a non zero primordial velocity dispersion for WDM particles.  The streaming
velocities were generated using the corresponding expression for the average
primordial velocity $\langle
v(z)\rangle$ given e.g. in Section~4 in~\citet{Boyarsky:08c}:
\begin{equation}
  \label{eq:vnrp}
  \langle v(z)\rangle = F_\wdm\frac{3.151 T_\nu(z)}{m_\wdm} 
\end{equation}
or numerically:
\begin{equation}
\langle v(z)\rangle= 15.7 \times F_\wdm\left( \frac{1+z}{100} \right)
\left( \frac{1~\kev}{m_\wdm} \right) ~\mathrm{km.s^{-1}}.
\end{equation}
where $z$ is the redshift of the beginning of simulation, $m_\wdm$ is the mass
of the WDM particle. We had kept the amplitude of primordial velocity constant
and equal to~(\ref{eq:vnrp}) and randomly chosen the direction of velocities
of the particles to add them to the Zel'dovich velocities, generated with the
GRAFIC2 package.

\section{Numerical Simulations} 
\label{sec:sims}

All simulations have been performed with {\sc pkdgrav}, a tree code written by
Joachim Stadel and Thomas Quinn (Stadel 2001). The code uses spline kernel
softening, for which the forces become completely Newtonian at 2 softening
lengths.  Individual time steps for each particle are chosen proportional to
the square root of the softening length, $\epsilon$, over the acceleration,
$a$: $\Delta t_i = \eta\sqrt{\epsilon/a_i}$. Throughout, we set $\eta = 0.2$,
and we keep the value of the softening length constant in comoving coordinates
during each run. The physical values of $\epsilon$ at $z=0$ are listed in
Table~\ref{tab:sims}.  Forces are computed using terms up to hexadecapole
order and a node-opening angle $\theta$ which we change from $0.55$ initially
to $0.7$ at $z=2$.  This allows a higher force accuracy when the mass
distribution is nearly smooth and the relative force errors can be large.

\begin{table}
\centering
  \begin{tabular}{lcccr}
    \hline  Name &  Box  size  & N  &  part. mass  &  force  soft.\\
    & $[h^{-1} {\rm Mpc}]$ &  & $[h^{-1}M_{\odot}]$  & $[h^{-1}{\rm kpc}]$  \\
    \hline\\
    $\Lambda$CDM--20   & 20   & $350^3$ & 1.31e7  & 1.42 \\ 
    $\Lambda$CDM-45   & 45   & $350^3$ & 1.49e8 & 3.21 \\ 
    $\Lambda$CDM-90   & 90   & $400^3$ & 7.99e8 &  5.62\\ 
    $\Lambda$CDM-180  & 180  & $400^3$ & 6.39e9 & 11.25\\ 
    WDM-20  & 20   & $350^3$ & 1.31e7  & 1.42 \\ 
    WDM-45  & 45   & $350^3$ & 1.49e8  & 3.21 \\ 
    m1-20vel& 20   & $350^3$ & 1.31e7  & 1.42 \\ 
    m1-20   & 20   & $350^3$ & 1.31e7  & 1.42 \\ 
    m1-45   & 45   & $350^3$ & 1.49e8 & 3.21 \\ 
    m1-90   & 90   & $400^3$ & 7.99e8 & 5.62 \\ 
    m1-180  & 180  & $400^3$ & 6.39e9 & 11.25\\   
    m5-20vel& 20   & $350^3$ & 1.31e7  & 1.42 \\ 
    m5-20   & 20   & $350^3$ & 1.31e7  & 1.42 \\ 
    m5-45   & 45   & $350^3$ & 1.49e8 & 3.21 \\ 
    m5-90   & 90   & $400^3$ & 7.99e8 &  5.62\\ 
    m5-180  & 180  & $400^3$ & 6.39e9 & 11.25\\ 
    \hline 
\end{tabular}
  \caption{N-body simulation parameters}
\label{tab:sims}
\end{table}

Table~\ref{tab:sims} lists all of the simulations used in this work.  We have
run simulations for several different box sizes, which allows us to probe halo
masses covering the entire range $10^{10} \hMsun \le M \le 10^{14} \hMsun$.
For the small box simulations ($L=20$ Mpc/h) we have, for each CWDM model, two
different runs, with and without thermal velocities.

\begin{figure}
  \includegraphics[width=8cm,angle=270]{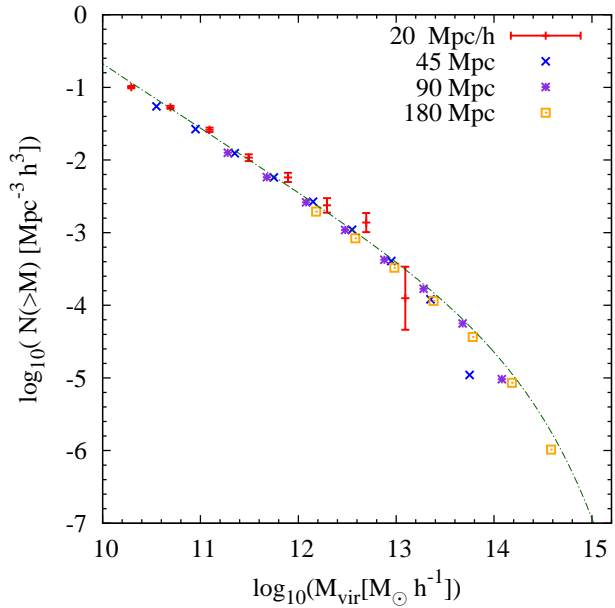}
  \caption{Halo mass function for all three models at $z=0$ follow predictions
    of $\Lambda$CDM with the same cosmological parameters (green dashed line
    is the \citet{Warren:06} predictions for $\Lambda$CDM). Shown are the
    points for \m5 model, the other two models (\m1 and WDM) give very similar
    results and are not shown.}
\label{fig:mf}
\end{figure}

Fig. \ref{fig:mf} shows the halo mass function for the \m5 model at $z=0$.
Only haloes with viral masses larger than $ 2\times10^{10} \hMsun$) are shown,
and different symbols (and colors) refer to different
box size simulations.  The green line is the \citet{Warren:06} prediction
for a pure $\Lambda$CDM model with the same cosmological
parameters as our CWDM models.

\noindent
The \citet{Warren:06} prediction is a good representation of our data at high
and low masses. The \m1 and WDM models present similar behaviours as the \m5 one.

\subsection{Halo parameters}
\label{ssec:param}

In all of the simulations, dark matter haloes are identified using a spherical
overdensity (SO) algorithm. We use a time varying virial density contrast
determined using the fitting formula presented in \citet{Mainini:03b}.  We
include in the halo catalog all the haloes with more than 500 particles
(see \cite{Maccio:08} for further details on our halo finding algorithm).
For each SO halo in our sample we determine a set of parameters, including the
virial mass and radius, the concentration parameter, the angular momentum, the
spin parameter and axis ratios (shape).  Below we briefly describe how these
parameters are defined and determined. A more detailed discussion can be found
in \citet{Maccio:07,Maccio:08}.  Finally following \citet{Maccio:07}, we split
our halo sample into unrelaxed and relaxed haloes. In the rest of the paper we
will only discuss the properties of relaxed haloes.

\subsubsection{Concentration parameter}

\begin{figure} \centering
  \includegraphics[width=\linewidth]{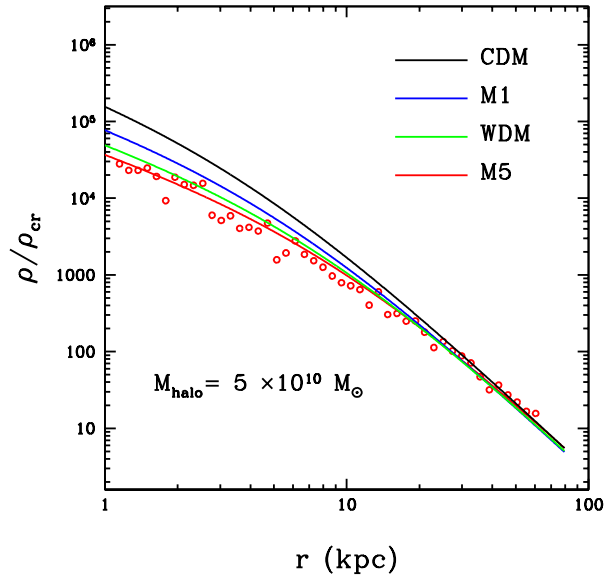}
  \caption{Density profiles for CWDM and WDM models. The plot shows the change
    in the inner slope of the halo density profile for one of the haloes in
    the simulation with the box size $L=20$~Mph/h (as compared to the CDM --
    black line).  Points and red line (the NFW fit) are for the \m5 model.
    For \m1 and WDM models we show only the NFW fit (to make the plot less
    crowded).  Black line shows the predicted profile for $\Lambda$CDM using
    the fitting formula from~\protect\citet{MunozCuartas:2010ig}.}
\label{fig:profile}
\end{figure}

To compute the concentration of a halo we first determine its density
profile. The halo centre is defined as the location of the most bound halo
particle, and we compute the density ($\rho_i$) in 50 spherical shells, spaced
equally in logarithmic radius. Errors on the density are computed from the
Poisson noise due to the finite number of particles in each mass shell.  The
resulting density profile is fit with a NFW profile \citep{Navarro:96}~
\footnote{In this paper we do not resolve (due to resolution) 
and do not discuss the inner slope of DM density profiles, as 
for example in \citet{Maccio:2012}. 
Therefore, possible deviations from the NFW profile due to WDM effects, 
even appearance of a core, are not considered.}:
\begin{equation}
\frac{\rho(r)}{\rhocrit} = \frac{\delta_{\rm c}}{(r/\rs)(1+r/\rs)^2},
\label{eq:nfw}
\end{equation}
During the  fitting procedure  we treat both  $\rs$ and  $\delta_c$ as
free  parameters.   Their values,  and  associated uncertainties,  are
obtained via a $\chi^2$  minimization procedure using the Levenberg \&
Marquardt method.  We define the r.m.s. of the fit as:
\begin{equation}
\rhorms = \frac{1}{N}\sum_i^N { (\ln \rho_i - \ln \rho_{\rm m})^2}
\label{eq:rms}
\end{equation}
where $\rho_{\rm m}$ is the fitted NFW density distribution.
Finally, we define the concentration of the halo, $\cvir \equiv\Rvir/\rs$,
using the virial radius obtained from the SO algorithm, and we define the
error on $\log c$ as $(\sigma_{\rs}/\rs)/\ln(10)$, where $\sigma_{\rs}$ is the
fitting uncertainty on $\rs$.

\subsubsection{Shape parameter}

Determining the shape of a three-dimensional distribution of particles is a
non-trivial task (e.g., \citealt{Jing:02}).  Following \citet{Allgood:06}, we
determine the shapes of our haloes starting from the inertia tensor.  As a
first step, we compute the halo's $3 \times 3$ inertia tensor using all the
particles within the virial radius.  Next, we diagonalize the inertia tensor
and rotate the particle distribution according to the eigenvectors.  In this
new frame (in which the moment of inertia tensor is diagonal) the ratios
$s=a_3/a_1$ and $p=a_2/a_1$ (where $a_1 \geq a_2 \geq a_3$) are given by:
\begin{equation}
s \equiv {a_3 \over a_1} = \sqrt{ { \sum m_i z_i^2} \over \sum { m_i x_i^2}}\\
p \equiv {a_2 \over a_1} = \sqrt{ { \sum m_i z_i^2} \over \sum { m_i y_i^2}}.
\end{equation}

Next we again compute the inertia tensor, but this time only using the
particles  inside the  ellipsoid defined  by $a_1$,  $a_2$, and  $a_3$.
When deforming the ellipsoidal volume of the halo, we keep the longest
axis  ($a_1$) equal  to the  original radius  of the  spherical volume
($\Rvir$).  We  iterate this procedure  until we converge to  a stable
set of axis ratios.

\section{The concentration mass relation}
\label{sec:results}

\begin{figure} \centering
  \includegraphics[width=\linewidth]{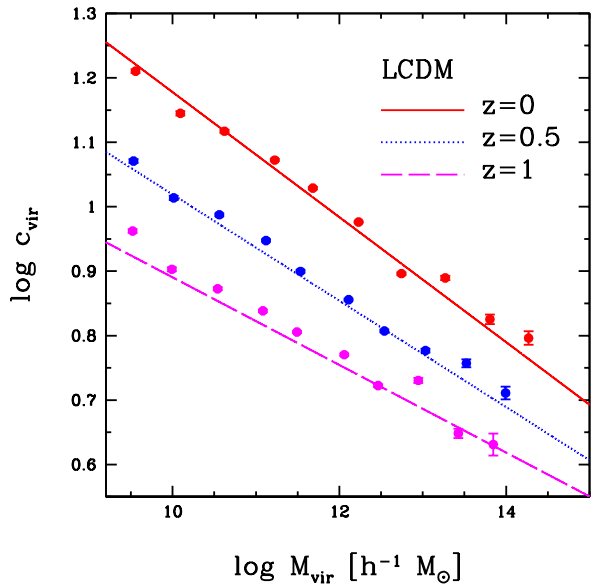}
  \caption{Mass and redshift dependence of the concentration parameter for the
$\Lambda$CDM model with the same parameters as the mixed ones.  Points
    with error-bars are simulation results, the straight lines are linear fits (in a 
log-log space) to the simulation results.}
\label{fig:conclcdm}
\end{figure}

In Figure \ref{fig:conclcdm}, we show the concentration mass relation for relaxed haloes
in the $\Lambda$CDM model. In our mass range the $c_{\rm vir}-M_{\rm vir}$ relation is well fitted
by a single power law at all redshifts, in agreement with several previous results (e.g. 
\cite{Klypin:10}, and references therein). 
The best fitting power law can be written as:
\begin{equation}
\log(c) = a(z) \log(M_{\rm vir}/[h^{-1} \Msun])+b(z).
\end{equation}
The fitting parameter $a(z)$ and $b(z)$ are function of redshifts and are reported
in table \ref{tab:param}. These parameters are very not far from the ones suggested
in \citet{MunozCuartas:2010ig}. From now on we will use 
these linear fits to compare (C)WDM models and the standard $\Lambda$CDM one.

\begin{table}
\centering
  \begin{tabular}{lccc}
    \hline  Name &  $z=0$  & $z=0.5$  &  $z=1$ \\
    \hline
    a(z)  & -0.0973  & -0.0828  & -0.0684\\ 
    b(z)  & 2.157    & 1.845    & 1.578 \\ 
    $\alpha(z)$  & -4.78e-6    &  -5.3095e-6   & -4.6561e-6  \\ 
    $\beta(z)$  & 0.766 & 0.729  & 0.681 \\ 
    \hline 
\end{tabular}
  \caption{Fitting parameter for the concentration and shape mass dependence}
\label{tab:param}
\end{table}

\begin{figure*}
  \includegraphics[width=8.5cm]{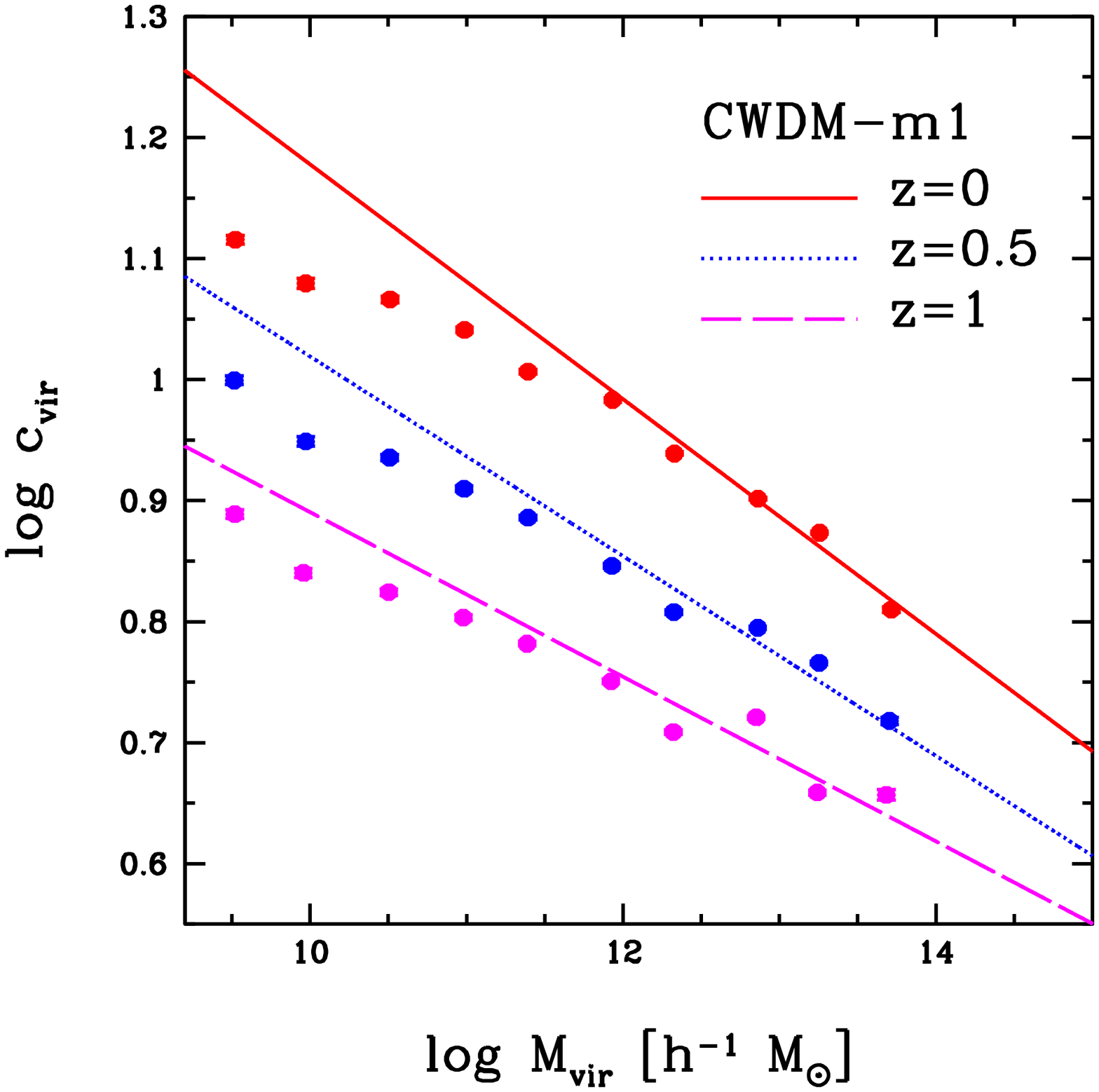}~ 
  \includegraphics[width=8.5cm]{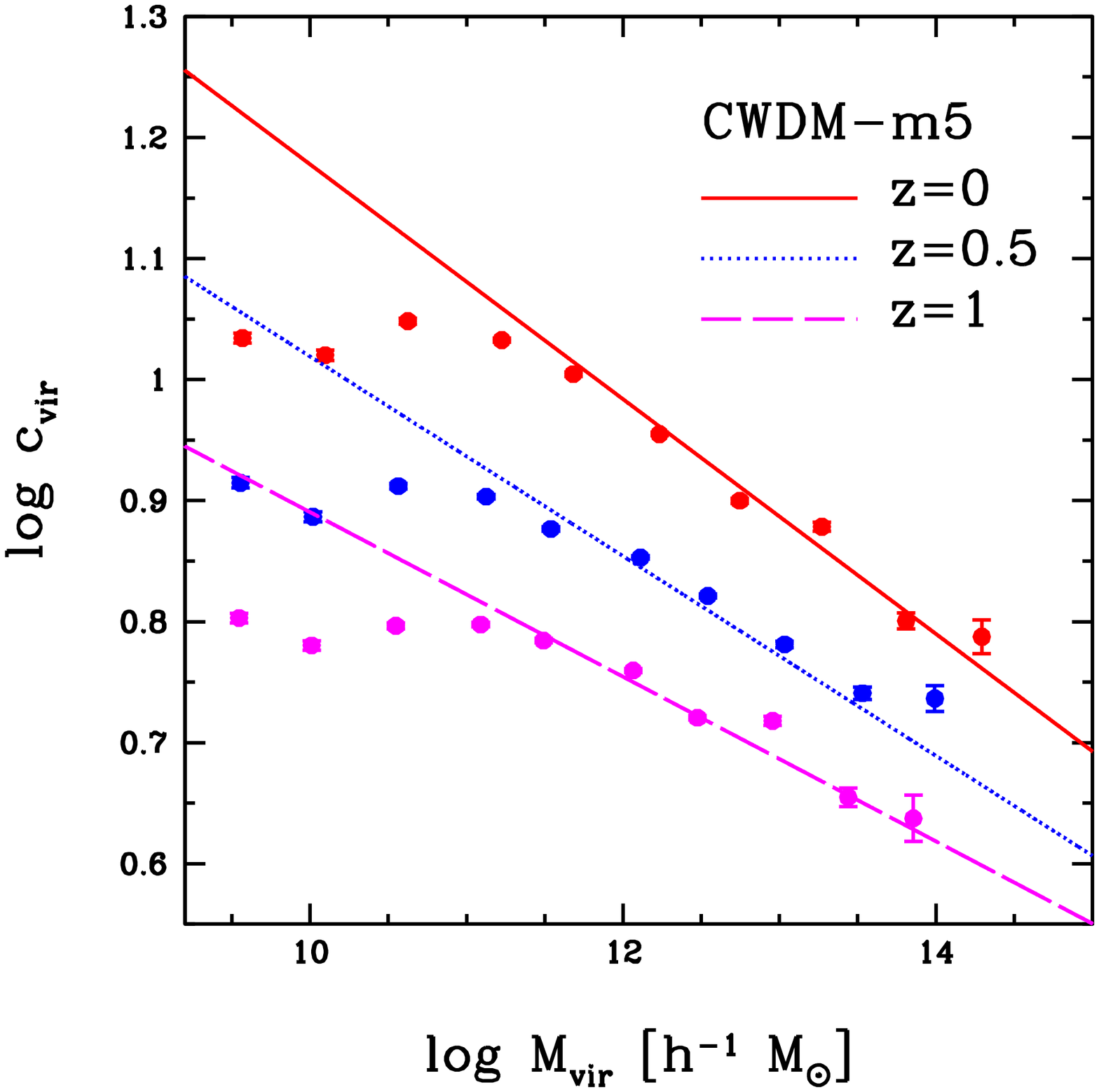}
 \caption{Mass and redshift dependence of the concentration parameter. Points
    with error-bars are CWDM simulation (\m1 and \m5 models at left and right
    panel correspondingly) results, 
    the three straight lines are the fits to the ΛCDM results from figure 4.
    From top to
    bottom: $z=0$, $z=0.5$, $z=1$. One sees that \m1 model is ``colder''
    (closer to the CDM) than \m5 at all redshifts. }
  \label{fig:Cm1}
\end{figure*}

Figure \ref{fig:Cm1} (right) shows the mass and redshift dependence of the
concentration parameter for the \m5 model. Results for $z=0$, $z=0.5$ and $z=1$
are shown from top to bottom. The error-bars on the points show the error in
the mean concentration value (while the scatter is of the order of 0.3 dex independently on the mass scale)
The three lines are the fits to the $\Lambda$CDM results from figure \ref{fig:conclcdm}.

At high masses the CWDM models basically agree with the pure cold dark matter predictions.
The situation is different for masses below $10^{12} \hMsun$, where the CWDM model predicts lower concentrations
with respect to the $\Lambda$CDM. For our lowest mass bin ($M\approx 10^{10} \hMsun$) the concentration is lower by 40\%
and there is a clear indication of flattening of the concentration mass relation, with an almost flat 
relation for $\log(M)<10.5$. This flat tail of the cM relation at low masses is already in place at $z=1$ 
and the difference between CDM and CWDM appears to be redshift independent.

This results is not surprising. In CWDM models the formation of small haloes
is delayed with respect to CDM due to the lack of power on small scales. The
concentration parameter is related to the density of the universe at the time
of the halo formation \citep{Wechsler:02}; since the density of the universe
decreases with time, a later formation time implies a lower value for the
concentration.

The \m1 model behaves similarly to \m5 (left panel of figure \ref{fig:Cm1}), 
showing a lower values for the concentration parameter (with respect to $\Lambda$CDM) at small
masses. In this case the difference with $\Lambda$CDM is less
pronounced and at the lowest mass scales ($M\approx 10^{11} \hMsun$) it is
less than 15\%.

\begin{figure}
  \includegraphics[width=\linewidth]{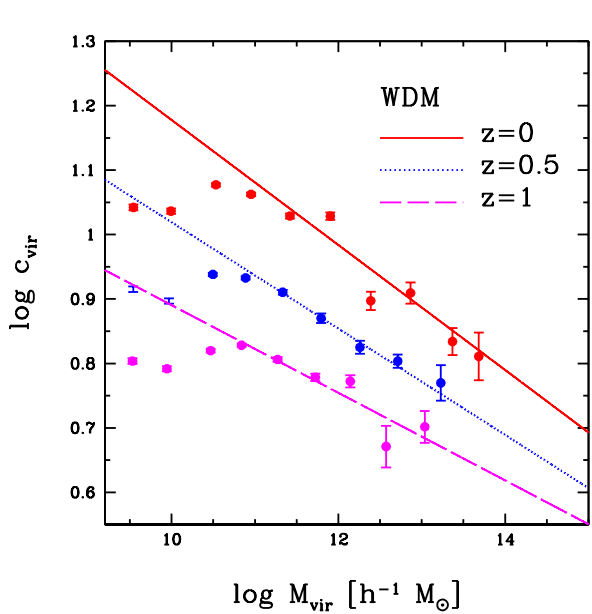}
 \caption{Same as fig. \ref{fig:Cm1} for the WDM simulation.
   Notice the similarity with Fig.  \ref{fig:Cm1}, right panel}
  \label{fig:Cmwdm}
\end{figure}

The difference between \m1 and \m5 is a direct consequence of
the different power spectrum at small scales.  As shown in Figure \ref{fig:tf}
the \m5 model has less power at scales $k > 10$ h/Mpc, which correspond
to mass scales of the order $8 \times 10^{10} \hMsun$. This lower power
results in a later formation time for these halos and hence lower concentration.
This is also confirmed by fig \ref{fig:Cmwdm}, where the concentration-mass 
relation for the WDM simulation is presented. Results for the pure WDM model are very 
similar to the \m5 model, confirming the relation between initial power spectrum 
(fig \ref{fig:tf}) and halo concentration.

The difference between the two CWDM models shows that even models with the 
same free-streaming length (as \m1 and \m5) could lead to different
halo internal structures. As a consequence 
the characterization of any models with a {\it warm} component only through 
its matter power spectrum suppression could lead to misleading results, since this 
single parameter is not capable to fully describe the effects
of a warm dark matter candidate on the structure formation process.

\subsection{Effects of thermal velocities}
\label{sec:effects-velocities}

For the small boxes (20 Mpc/h) we also run an additional simulation that
included a thermal velocity component in the initial conditions.  This thermal
component is few percent of the initial velocity due to the potential field
(according to the Zel'dovich approximation), nevertheless it is important to
test its effects (if any) on the halo internal structure.

\begin{figure}
  \includegraphics[width=8.0cm,angle=270]{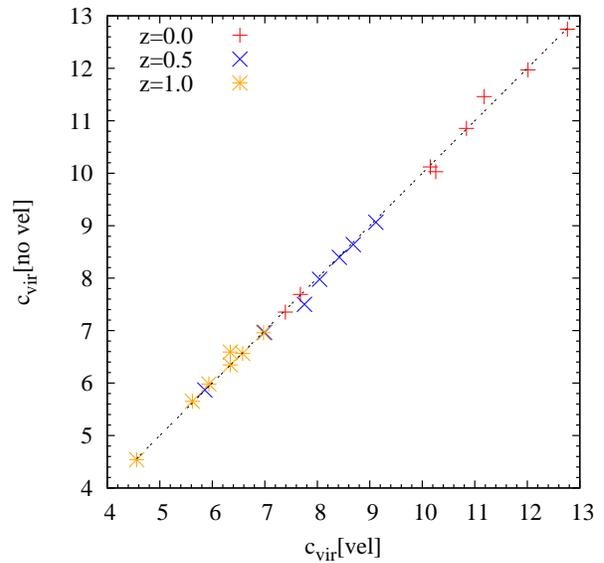}
  \caption{Effects of thermal velocities on the
    concentration parameter in the 20 $h^{-1}$ Mpc box.  Different symbols
    refer to different redshifts.}
  \label{fig:velnovel}
\end{figure}

Figure \ref{fig:velnovel} shows the 
one-to-one comparison of the median $c_{vir}$ 
at a given halo mass for the model \m1, 
with and without primordial thermal velocities. 
As expected given the magnitude of the
thermal velocity component, the effect is negligible.  The model \m5 presents the
same behaviour.
The thermal velocity component will play a role on smaller scales (for example
for Milky Way satellites) and we plan to address this issue in a forthcoming
paper.

\subsection{Effects on halo shape}
\label{sec:effects-halo-shape}

There is a well known relation between halo shape and mass with low mass
haloes being less triaxial (higher value for the $s$ parameter) than high mass ones
\citep{Allgood:06,Maccio:08,MunozCuartas:2010ig}.  
This relation is usually explained by assuming that the halo ``triaxiality'' correlates with formation
time, such that early halos have more time to virialize and hence reach a more
equilibrated, less triaxial configuration.  In this is the case we could possibly see 
a different trend between halo shape and mass in the CWDM
simulation with respect to the $\Lambda$CDM.

\begin{figure} \centering
  \includegraphics[width=\linewidth]{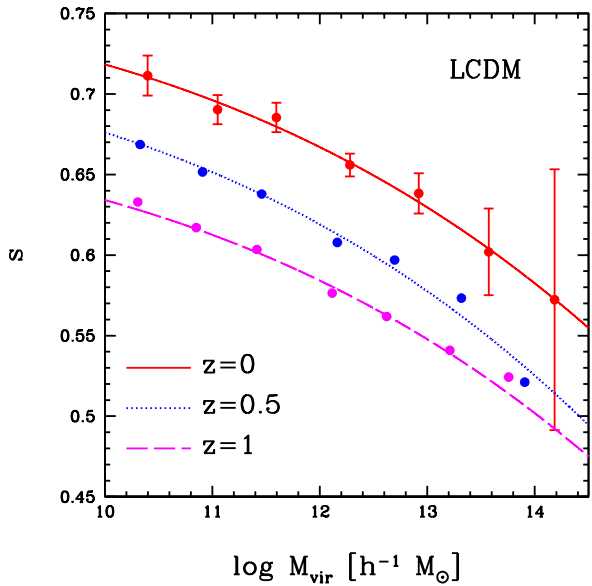}
  \caption{Mass and redshift dependence of the halo shape for the
$\Lambda$CDM model with the same parameters as the mixed ones.  Points with error-bars (shown only for
the $z=0$ case) are simulation results, lines are polynomial fits to Eq. (9). }
\label{fig:shapelcdm}
\end{figure}

Figure \ref{fig:shapelcdm} shows the redshift and mass dependence of the shape parameter $s$ for the $\Lambda$CDM
model. In order to fit this relation we used the same equations as suggested by \citet{MunozCuartas:2010ig}:
\begin{equation}
s(z,M) = \alpha(z)(\log(M_{\rm vir})/[h^{-1} \Msun])^4+\beta(z).
\end{equation}
\label{eq:shape}
The values of the fit parameters ($\alpha(z),\beta(z)$) are listed in Table \ref{tab:param}.

\begin{figure*}
  \includegraphics[width=8.0cm]{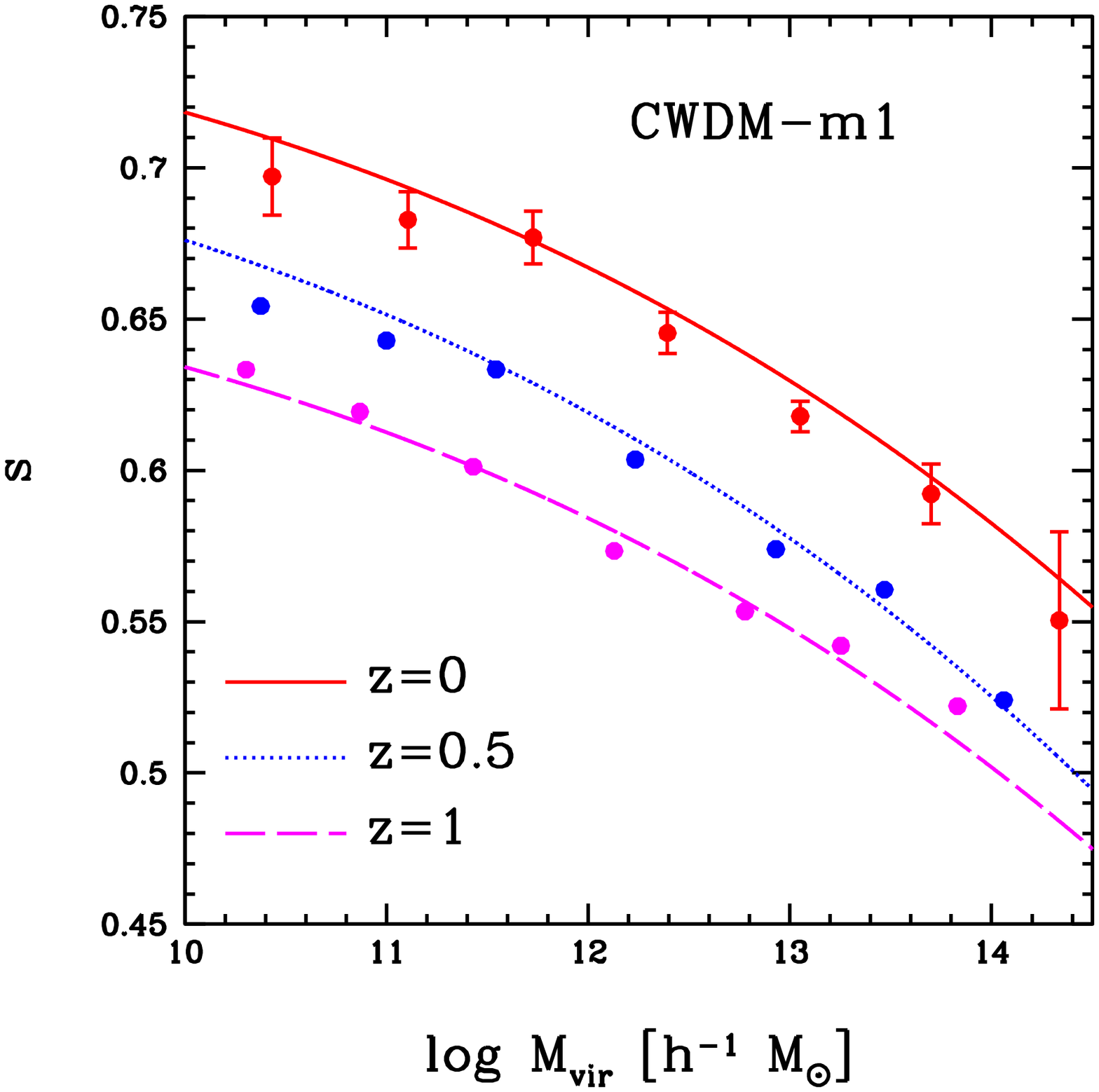}~%
  \includegraphics[width=8.0cm]{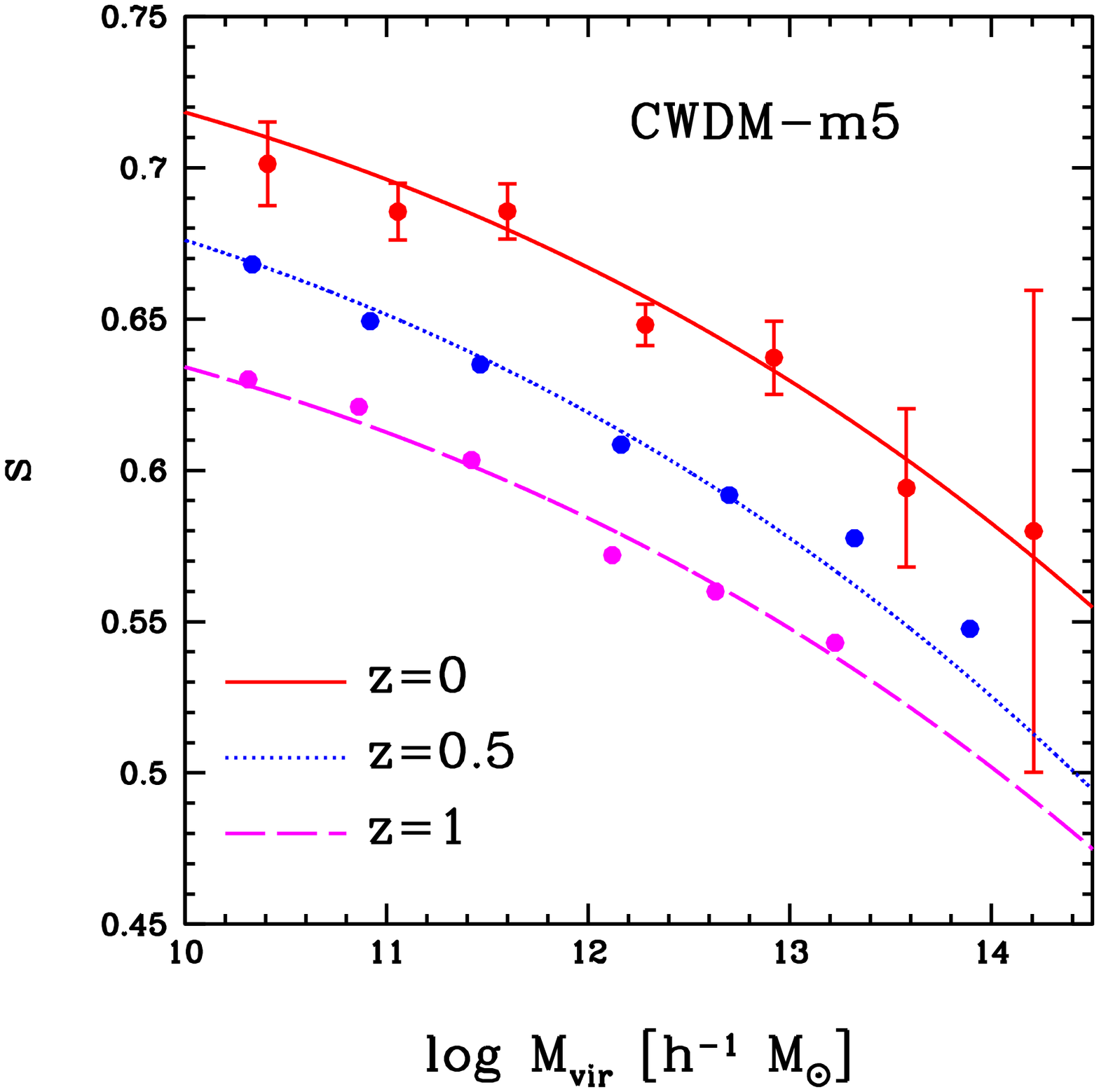}
\caption{Mass and time evolution of the shape of dark matter haloes quantified
  via the $s\equiv a3/a1$ ratio.  Lines are polynomial fits to Eq. (9) as in 
Fig. 8. Left panel -- \m1 model,
  right panel -- \m5 model}
\label{fig:shape.m1}
\end{figure*}

Figure \ref{fig:shape.m1} shows the redshift evolution of the halo shape,
quantified via the minor to major axis ratio: $s\equiv a3/a1$, for the \m1
and \m5 models (left and right panels in Fig.~\ref{fig:shape.m1}
correspondingly).

In this case the CWDM models behaviour is very similar to $\Lambda$CDM and the
two family models ($\Lambda$CDM and CWDM) seem to be consistent within the errors.
Nevertheless there is a small hint for a lower values for the $s$ parameter 
(hence a larger triaxiality)  at the very low mass bins ($\approx 10^{10.3} \hMsun$), as expected
due to the later formation times at these mass scales.


\section{Summary and discussion}
\label{sec:summ}

Interest in warm dark matter models has increased in the last years, mainly
motivated by the possible need for cored (or very low concentration) dark
matter halo density profiles for dwarf galaxies in the local group \citep[see
e.g][]{Walker:11,Amorisco:11}, or by solution of various CDM ``over-abundance
problems''~\citep{Klypin:99,Moore:99b}, see
e.g. \citep{Maccio:09,Strigari:10,BoylanKolchin:11,Papastergis:2011xe,Lovell:11}.

While it is still under debate what is the effects of baryons on the dark
matter distribution on such small scales (e.g. \citealt{Pontzen:11,Maccio:11}
and references therein); it is worth studying through numerical simulations
the effects of a warm component on the inner structure of DM haloes.  It had
been repeatedly argued that ``pure warm'' dark matter models (with a
cutoff-like suppression of the matter power spectrum) could not be a solution
to the core-cusp problem as the required free-streaming would be in a stark
contradiction with Lyman-$\alpha$ bounds~\citep[see
e.g.][]{Strigari:06,deNaray:2009xj}.

In this paper we have performed the first N-body simulations for {\it cold
  plus warm} dark matter models (CWDM). In this class of models the dark
matter particle candidate (e.g. sterile neutrinos, gravitino, axions) can be
produced via two co-existing mechanisms and therefore the primordial spectrum
is a superposition of a cold and a warm component, plus a complicated
non-thermal velocity spectrum (e.g. \citealt{Boyarsky:08d})

We have mainly focused our attention on the inner structure of dark matter
haloes in CWDM models, and its evolution with redshift.  We have adopted the
commonly used {\it concentration} parameter (e.g. \citealt{Maccio:08}) to
parameterize the modification in CWDM models with respect to the standard Cold
Dark Matter (CDM) model.

We show that in our models the number of halos remains the same as in
$\Lambda$CDM down to the smallest scales resolved in the simulations
($\sim10^{10}M_\odot$). At the same time, keeping the free-steaming scale the
same, the variation of WDM fraction is able to reduce the concentration
parameter on mass scales as high as $M \approx 10^{12} \hMsun$, two/three
order of magnitude above the free streaming mass of the model.  On dwarf
galaxies mass scale($M\approx 10^{10} \hMsun$) the concentration parameter is
almost half of the value predicted by $\Lambda$CDM and we see a clear sign of a
strong flattening of the concentration mass relation.  As a consequence we
expect an even stronger concentration reduction on lower mass scales (below
our resolution limits), mass scales directly explored by the Milky Way
satellites. This decrease of the concentration of such large halos may explain
why the ``cuspy'' matter distributions are not supported by observations of
the rotation curves of spiral galaxies~(see
e.g.~\citealt{Salucci:00,Oh:08,Spano:07,deNaray:2009xj}), thus resolving one
of the major challenges for the CDM cosmological model.

Another interesting result is the intrinsic difference in the concentration 
mass relation between the \m1 and \m5 models, that clearly shows
that even models with the same free-streaming length could lead to different
halo internal structures. As a consequence this calls for a detailed analysis
of {\it any} model with a warm component, since a single parameter description
through the free-streaming length  is not capable to fully capture
the effects on structure formation.

In~\citet{Maccio:2012} it was shown that the primordial velocities required to
produce the cores of observable sizes in pure WDM cosmologies are so large
that too few galaxies would be formed. This is because in pure warm dark
matter models the phase-space density, defining the size of the core and
suppression in the halo number density are defined by the same parameter --
the average velocity of particles. Our results show that in the CWDM models
the presence of the second parameter, $F_\wdm$, allows to control the
``warmness'' of dark matter, separating the scales of modification of the
matter power spectrum (halo number density) and that of modification of the
density profile (size of the core).

We stress that unlike the recent work of~\citet{Maccio:2012} this effect is
not related to the finite phase-space density (the phase-space density of
particles in our simulations is the same as in pure CDM ones).  We ascribe the
lower concentration at small masses to a shift in the halo formation time,
that results in a delayed formation for CWDM models with respect to CDM, as
already found in previous studies \citep{Eke:2000av,Lovell:11}.  This delayed
formation time could also be responsible of the lower ratio between the halo minor and major
axis ($s$ parameter) that we saw at very low masses (even if the
CWDM models are consistent with $\Lambda$CDM within the error-bars).
We did not find any effect on the halo
spin parameter distribution, which looks almost identical in all models.

We also tested the effect of explicitly including thermal velocities in the
N-body initial conditions, We did not see any visible change in the
concentration mass relation on the mass scales we are able to resolve. We plan
to study in more details the importance of this thermal velocity component in
a forthcoming paper, by employing ``zoomed'' high resolution simulations of
single dark matter haloes.

Along with the Lyman-$\alpha$ forest method, the weak lensing surveys can be
used to probe further clustering properties of dark matter particles as
sub-galactic scales, as the next generation of these surveys (such as
e.g. KiDS, LSST, WFIRST, Euclid) will be able to measure the matter power
spectrum at scales down to $1-10$~h/Mpc with a few percent
accuracy. \citet{Markovic:10,Smith:11} argued that the next generation of
lensing surveys can provide sensitivity, compatible with the existing
Lyman-$\alpha$ bounds \citep{Viel:06,Seljak:06,Boyarsky:08c}).  As in the case
of the Lyman-$\alpha$ forest method the main challenge for the weak lensing is
to properly take into account baryonic effects on matter power spectrum. The
suppression of power spectrum due to primordial dark matter velocities can be
extremely challenging to disentangle from the modification of the matter power
spectrum due to baryonic feedback~\citep{Semboloni:11,vanDaalen:11}.  Finally,
the modified concentration mass relation can be probed with the weak lensing
surveys~\citep[see e.g.][]{Mandelbaum:08,King:11} if their sensitivity can be
pushed to halo masses below roughly $10^{12}M_\odot$.

While the observational difference between pure cold and mixed cosmologies is
not drastic, clarifying this issue would have profound impact on the
fundamental physics questions.  Cold+Warm dark matter models are still in an
infant state and an effort comparable with that, invested in theoretical
investigation of structure formation of pure CDM may be needed before this
question will be finally settled.

\section*{Acknowledgments}

We thank the anonymous referee whose comments have strongly improved the presentation
of this paper.
All numerical simulations were performed on the Theo and on PanStarrs2
clusters of the Max-Planck-Institut f\"ur Astronomie at the Rechenzentrum in
Garching.


\let\jnlstyle=\rm\def\jref#1{{\jnlstyle#1}}\def\aj{\jref{AJ}}
  \def\araa{\jref{ARA\&A}} \def\apj{\jref{ApJ}\ } \def\apjl{\jref{ApJ}\ }
  \def\apjs{\jref{ApJS}} \def\ao{\jref{Appl.~Opt.}} \def\apss{\jref{Ap\&SS}}
  \def\aap{\jref{A\&A}} \def\aapr{\jref{A\&A~Rev.}} \def\aaps{\jref{A\&AS}}
  \def\azh{\jref{AZh}} \def\baas{\jref{BAAS}} \def\jrasc{\jref{JRASC}}
  \def\memras{\jref{MmRAS}} \def\mnras{\jref{MNRAS}\ }
  \def\pra{\jref{Phys.~Rev.~A}\ } \def\prb{\jref{Phys.~Rev.~B}\ }
  \def\prc{\jref{Phys.~Rev.~C}\ } \def\prd{\jref{Phys.~Rev.~D}\ }
  \def\pre{\jref{Phys.~Rev.~E}} \def\prl{\jref{Phys.~Rev.~Lett.}}
  \def\pasp{\jref{PASP}} \def\pasj{\jref{PASJ}} \def\qjras{\jref{QJRAS}}
  \def\skytel{\jref{S\&T}} \def\solphys{\jref{Sol.~Phys.}}
  \def\sovast{\jref{Soviet~Ast.}} \def\ssr{\jref{Space~Sci.~Rev.}}
  \def\zap{\jref{ZAp}} \def\nat{\jref{Nature}\ } \def\iaucirc{\jref{IAU~Circ.}}
  \def\aplett{\jref{Astrophys.~Lett.}}
  \def\apspr{\jref{Astrophys.~Space~Phys.~Res.}}
  \def\bain{\jref{Bull.~Astron.~Inst.~Netherlands}}
  \def\fcp{\jref{Fund.~Cosmic~Phys.}} \def\gca{\jref{Geochim.~Cosmochim.~Acta}}
  \def\grl{\jref{Geophys.~Res.~Lett.}} \def\jcp{\jref{J.~Chem.~Phys.}}
  \def\jgr{\jref{J.~Geophys.~Res.}}
  \def\jqsrt{\jref{J.~Quant.~Spec.~Radiat.~Transf.}}
  \def\memsai{\jref{Mem.~Soc.~Astron.~Italiana}}
  \def\nphysa{\jref{Nucl.~Phys.~A}} \def\physrep{\jref{Phys.~Rep.}}
  \def\physscr{\jref{Phys.~Scr}} \def\planss{\jref{Planet.~Space~Sci.}}
  \def\procspie{\jref{Proc.~SPIE}} \let\astap=\aap \let\apjlett=\apjl
  \let\apjsupp=\apjs \let\applopt=\ao

\label{lastpage}
\end{document}